\documentstyle[11pt,epsfig,amsbsy]{article}
\def\be{\begin{equation}}
\def\ee{\end{equation}}
\def\ba{\begin{eqnarray}}
\def\ea{\end{eqnarray}}

\def\ga{\mathrel{\raise.3ex\hbox{$>$\kern-.75em\lower1ex\hbox{$\sim$}}}}
\def\la{\mathrel{\raise.3ex\hbox{$<$\kern-.75em\lower1ex\hbox{$\sim$}}}}

\newcommand{\fr}[2]{\frac{#1}{#2}}

\newcommand{\PRD}{Phys.\ Rev.\ D}
\newcommand{\MNRAS}{Mon.\ Not.\ R.\ Astron.\ Soc.}

\newcommand{\ApJ}{Astrophys.\ J.\ }

\newcommand{\omde}{\omega_{\rm{de}}}
\newcommand{\Omo}{\Omega_{\rm{m}}^{0}}
\newcommand{\Odeo}{\Omega_{\rm{de}}^{0}}
\newcommand{\Oma}{\Omega_{\rm{m}}(a)}
\newcommand{\OLo}{\Omega_{\Lambda}^{0}}
\newcommand{\OLa}{\Omega_{\Lambda}(a)}
\newcommand{\Odea}{\Omega_{\rm{de}}(a)}
\newcommand{\rde}{\rho_{\rm{de}}}
\newcommand{\rhom}{\rho_{\rm{m}}}
\newcommand{\rcr}{\rho_{\rm{cr}}}

\newcommand{\dm}{\delta_{\rm{m}}}
\newcommand{\kpc}{\rm{kpc}}
\newcommand{\Mpc}{\rm{Mpc}}
\newcommand{\sw}{\rm{sw}}
\newcommand{\ws}{\rm{ws}}
\newcommand{\cpt}{\rm{cpt}}
\newcommand{\bp}{\rm{bp}}
\newcommand{\lahav}{\rm{lahav}}
\newcommand{\step}{\rm{step}}
\newcommand{\arctanh}{\rm{arctanh}}
\begin{document}

\baselineskip=16pt
\begin{titlepage}
% \rightline{arXiv:0810.XXXX} \rightline{September 2008}
\begin{center}

\vspace{0.5cm}

\large {\bf Properties of the exact analytic solution of the growth factor and its applications}
\vspace*{5mm} \normalsize

{\bf Seokcheon Lee$^{\,1,2}$ and Kin-Wang Ng$^{\,1,2,3}$}

\smallskip
\medskip

$^1${\it Institute of Physics, Academia Sinica, \\
Taipei, Taiwan 11529, R.O.C.}

$^2${\it Leung Center for Cosmology and Particle Astrophysics, National Taiwan University, \\ Taipei, Taiwan 10617, R.O.C.}

$^3${\it Institute of Astronomy and Astrophysics, \\
 Academia Sinica, Taipei, Taiwan 11529, R.O.C.}

\smallskip
\end{center}

\vskip0.6in

\centerline{\large\bf Abstract}
There have been the approximate analytic solution \cite{Silveira} and several approximate analytic forms \cite{0508156,Carroll,0303112} of the growth factor $D_{g}$ for the general dark energy models with the constant values of its equation of state $\omde$ after Heath found the exact integral form of the solution of $D_{g}$ for the Universe including the cosmological constant or the curvature term. Recently, we obtained the exact analytic solutions of the growth factor for both $\omde = -1$ or $-\fr{1}{3}$ \cite{SK1} and the general dark energy models with the constant equation of state $\omde$ \cite{SK3} independently. We compare the exact analytic solution of $D_{g}$ with the other well known approximate solutions. We also prove that the analytic solutions for $\omde = -1$ or $-\fr{1}{3}$ in Ref. \cite{SK1} are the specific solutions of the exact solutions of the growth factor for general $\omde$ models in Ref. \cite{SK3} even though they look quite different. Comparison with the numerical solution obtained from the public code is done. We also investigate the possible extensions of the exact solution of $D_{g}$ to the time-varying $\omde$ for the comparison with observations.

\vspace*{2mm}
%\smallskip\newline

\end{titlepage}

%%%%%%%%%%%%%%%%%%%%%%%%%%%%%%%%%%%%%%%%%%%%%%%%%%%%%%%%%%%%%%%%%%%%%%%%%
\section{Introduction}
\setcounter{equation}{0}
%%%%%%%%%%%%%%%%%%%%%%%%%%%%%%%%%%%%%%%%%%%%%%%%%%%%%%%%%%%%%%%%%%%%%%%%%
The analysis of the luminosity distance as a function of redshift obtained from distant Type Ia supernovae discovered that the present Universe is expanding at an accelerating rate \cite{9805201,9812133,0611572}. One of the most popular solutions to this conundrum is introducing the so called ``dark energy'' (DE) which is the dominant energy contribution to the present energy of the Universe with its equation of state (EOS), $\omde < -\fr{1}{3}$ (for example, \cite{07054102}). The combined observations of the large scale structure (LSS) and of the cosmic microwave background (CMB) power spectra have confirmed the cosmic concordance ({\it i.e.} a flat universe with the present energy density contrast of the matter $\Omo \simeq 0.3$ and with that of the dark energy $\Odeo \simeq 0.7$) \cite{9505066,0407372,0412631,0501174,0608632,08030547}.

Due to our ignorance of the nature of the dark energy, it is practical to use the EOS of the DE, $\omde$ to characterize it \cite{0510017}. Moreover, $\omde$ is the quantity constrained by cosmological observations \cite{0403292,0411803}. Among the excess of models, the cosmological constant $\Lambda$ and a quintessence field are the most commonly proposed candidates for dark energy \cite{0207347}. The former is characterized by $\omde = -1$ and the latter is a dynamical scalar field leading to a time dependent EOS, $\omde (a)$. Also models with the constant $\omde = \rm{constant} \neq -1$ are important because the effects of the time varying $\omde(a)$ can be predicted by interpolating between models with constant $\omde$ \cite{9901388,0206372,0209093,0508156,08040413,08073140}.

The origin of the current accelerating Universe is still in dispute (see for example, \cite{08030982}). There are two major theories for this. One is the dark energy and the other is the modified theory of gravities (MG). However, MG are also able to be characterized by the effective EOS which is used for specifying DE models \cite{0303009,0305286}. Unfortunately, observations only probe the cosmological evolution of $\omde$ in an indirect way and there might be some ambiguities to differentiate DE with a specific MG model \cite{0612452}. However, in most cases, while the two models give the same cosmic background expansion history $H(a)$, they predict different growth rates for cosmic LSS \cite{0305286,0207199,0404599,0507263,07103885,08060937}.

Thus, it is important to probe the accurate background expansion history of the Universe in order to constrain the EOS of the dark energy ({\it i.e.} its energy density, $\rde$) precisely \cite{0403292,0411803}. Furthermore, the evolution of the matter density perturbation $\dm$ also depends on $\omde$ \cite{0206372,CDS,9708069,9906174,0304212}. The formation of the LSS depends on the sound speed of the DE too \cite{CDS,9906174,0112438,0307100,0504017,0512135,0601333}. However, in general DE models including the quintessence, the sound speed of DE is close or equal to that of light and the DE is not able to cluster on the scales of galaxy clusters and below \cite{CDS,9906174}. Consequently, the DE only affects the matter power spectrum on large scales ($> 100 \Mpc$) \cite{0206372}. Usually, the LSS measurements probe scales $100 \kpc \sim 100 \Mpc$ and thus we may not need to worry about the effect of the growth of perturbation of DE when interpreting the LSS survey data.

In sub-horizon scales ($k \gg aH$), all the matter density perturbation modes $\delta_{\rm{m}}(\vec{k},a)$ grow uniformly because the dark energy do not cluster ($\Omega_{\rm{de}} \delta_{\rm{de}} \ll \Omega_{\rm{m}} \delta_{\rm{m}}$) and only the pressureless dark matter contributes to the gravitational potential. Thus, the effect of the existence of DE appears only through the Hubble parameter, $H(a)$ and one can use the linear growth factor $D(a)$, defined by $\delta_{\rm{m}}(\vec{k},a) \propto \delta_{H}(\vec{k}) D(a)$. From the growth factor, the growth index $f$ (sometimes it is called as ``growth rate'') is defined as $f = \fr{d \ln D(a)}{d \ln a} \equiv \Omega_{\rm{m}}(a)^{\gamma}$ \cite{Peebles1,Peebles2,Lahav}. In a flat universe, the growth factor is obtained in the integral form for the cosmological constant $\Lambda$ \cite{Heath}. This solution is widely used with the approximate analytic form \cite{Carroll}. This solution is even extended to the general dark energy models $\omde \neq -1$ \cite{0303112} by using the well known growth index parameter $\gamma$ (sometimes it is called as ``growth index'') given in the literature \cite{WS}. It is also known that the approximate analytic solution of $D(a)$ is obtained in the general dark energy models with the constant $\omde$ \cite{Silveira}. We have currently available data for $f(a)$ at various redshifts with the large degree of uncertainty though \cite{0608632,0112161,0212375,0407377,0612400,0612401,08021944}.

In what follows, we analyze in detail the recently obtained exact analytic solution of the growth factor $D(a)$ with the general constant $\omde$ dark energy in a flat universe \cite{SK1,SK2,SK3,SK4}. We note that the well known analytic solution of $D(a)$ in Ref. \cite{Silveira} is the approximate solution which shows the different behaviors of both $D(a)$ and $f(a)$ from the exact ones for some DE models. We do confirm that the exact analytic solutions of the growth factor with $\omde = -1$ and $-\fr{1}{3}$ obtained in Ref. \cite{SK1} are the specific solutions of the exact solution of $D(a)$ for general $\omde$ given in Ref. \cite{SK3} even though they look quite different. In Sec. 3, we compare the cosmological evolution of $D(a)$ obtained from the well known approximate analytic forms of it with those of the exact analytic solution $D(a)$. We also compare the values of $f(a)$ from these two solutions. We compare the exact sub-horizon solution values with the ones obtained from the full numerical values by using CMBFAST \cite{CMBFAST}. We investigate $D(a)$ and $f(a)$ with a specific parametrization of $\omde$ and its applications to observations in Sec. 4. We reach our conclusions in Sec. 5.

%%%%%%%%%%%%%%%%%%%%%%%%%%%%%%%%%%%%%%%%%%%%%%%%%%%%%%%%%%%%%%%%%%%%%%%%%
\section{Sub-horizon scale growth factor}
\setcounter{equation}{0}
%%%%%%%%%%%%%%%%%%%%%%%%%%%%%%%%%%%%%%%%%%%%%%%%%%%%%%%%%%%%%%%%%%%%%%%%%
We use the flat Friedmann-Robertson-Walker universe to probe the sub-horizon scale linear density perturbations of matter $\dm$ in the matter dominated epoch, \ba H^2 \equiv \Bigl(\fr{\dot{a}}{a}\Bigr)^2 &=& \fr{8\pi G}{3}(\rhom + \rde) = \fr{8 \pi G}{3} \rcr \, , \label{H} \\ 2 \fr{\ddot{a}}{a} + \Bigl(\fr{\dot{a}}{a}\Bigr)^2 &=& - 8 \pi G \omde \rde \, , \label{dotH} \ea where $\omde$ is the equation of state (EOS) of dark energy, $\rcr$ is the critical energy density, $\rhom$ and $\rde$ are the energy densities of the matter and the dark energy, respectively. We consider the constant $\omde$ and set the present scale factor $a_{0} = 1$. At sub-horizon scales ($k \gg aH$), all interesting modes of the matter density perturbation $\dm(\vec{k},a)$ grow uniformly as long as the dark energy do not cluster \cite{0206372,9708069,9906174}. It means that we only consider the matter perturbation in Poisson equation in this scale. Thus, the growth factor $D_{g}(a)$ is defined as \be \dm(\vec{k},a) \propto \delta_{H}(k) D_{g}(a) \, ,\label{D} \ee where $\delta_{H}(k)$ is the scalar amplitude at the horizon crossing generated during the cosmic inflation. The alternative definition of the $D_{g}$ are also commonly used (see for example, \cite{09065036}) \be \dm(\vec{k},a) \equiv \delta_{0}(k) D_{g}(a) \, , \label{alternativeD} \ee where $\delta_{0}(k)$ is the present density contrast. Then we obtain the evolution equation of $D(a)$ from the linear density perturbation equations \cite{Peebles1, Bonnor}, \be \fr{d^2 D}{da^2} + \Biggl( \fr{d \ln H}{d a} + \fr{3}{a} \Biggr) \fr{d D}{d a} - \fr{4 \pi G \rhom}{(aH)^2} D = 0 \, . \label{daD} \ee We use $D$ in Eq. (\ref{daD}) instead of $D_{g}$ because the general solution of Eq. (\ref{daD}) does not guarantee that $D$ is the growing mode solution. We are able to find the exact analytic solution of $D(a)$ for any value of the constant $\omde$ \cite{SK1,SK2,SK3}. After replacing new parameters $Y = Q a^{3 \omde}$ and $Q = \fr{\Omo}{\Odeo}$  in Eq. (\ref{daD}), we obtain \be Y \fr{d^2 D}{dY^2} + \Bigl[1 + \fr{1}{6 \omde} - \fr{1}{2(Y+1)} \Bigr] \fr{d D}{dY} - \Bigl[\fr{1}{6 \omde^2 Y} - \fr{1}{6 \omde^2 Y(Y+1)} \Bigr] D = 0 \, . \label{dYD} \ee We replace a trial solution $D (Y) = c Y^{\alpha} B(Y)$ into Eq (\ref{dYD}) to get \ba && Y (1 + Y) \fr{d^2B}{dY^2} + \Biggl[ \fr{3}{2} - \fr{1}{6 \omde} + \Bigl( 2 - \fr{1}{6 \omde} \Bigr) Y \Biggr] \fr{d B}{dY} + \Biggl[ \fr{(3 \omde +2)(\omde -1)}{12 \omde^2} \Biggr] B = 0 \, , \nonumber \\ && {\rm when} \,\,\,\, \alpha = \fr{1}{2} - \fr{1}{6 \omde} \,\, .  \label{dYB} \ea The above equation becomes the so called ``hypergeometric" equation when we replace $X = -Y$, which has the complete solution \cite{Morse} \be B(Y) = c_{1} F [\fr{1}{2} - \fr{1}{2\omde}, \fr{1}{2} + \fr{1}{3 \omde}, \fr{3}{2} - \fr{1}{6 \omde}, -Y] + c_{2} Y^{\fr{1-3\omde}{6 \omde}} F[-\fr{1}{3\omde}, \fr{1}{2 \omde}, \fr{1}{2} + \fr{1}{6 \omde}, -Y] \, , \label{B} \ee
where $F$ is the hypergeometric function. Thus, the exact analytic solution of the above equation (\ref{daD}) is \ba D(a) &=& c_{1} \Biggl( \fr{\Omo}{\Odeo} \Biggr)^{\fr{3 \omde -1}{6 \omde}} a^{\fr{3 \omde -1}{2}} F \Bigl[ \fr{1}{2} - \fr{1}{2\omde}, \fr{1}{2} + \fr{1}{3 \omde}, \fr{3}{2} - \fr{1}{6 \omde}, -\fr{\Omo}{\Odeo} a^{3\omde} \Bigr] \nonumber \\ && \, + \, c_{2} F \Bigl[ -\fr{1}{3\omde}, \fr{1}{2 \omde}, \fr{1}{2} + \fr{1}{6 \omde}, -\fr{\Omo}{\Odeo} a^{3\omde} \Bigr] \, . \label{Dask} \ea $D(a)$ in Eq. (\ref{Dask}) is just the general solution of the second order differential equation (\ref{daD}). Thus it does not have any physical meaning yet. It may represent the growing mode, the decaying mode or none of them before we choose the integral constants $c_{1}$ and $c_{2}$. If we want to have the correct growing mode solution from the above analytic solution, then this solution should follow the behavior of growing mode solution at an early epoch, say $a_{i} \simeq 0.1$. In other words, the coefficients of the general solution should be fixed by using the initial conditions of the growth factor, \be D_{g}(a_i) \simeq a_{i} \hspace{0.2in} {\rm and} \hspace{0.2in} \fr{d D_{g}}{da} \Bigl|_{a_{i}} \simeq 1 \, . \label{inig} \ee After we fix the coefficients from the initial conditions, we are able to determine the growth factor $D_{g}(a)$ from the general form of solution $D(a)$ in Eq. (\ref{Dask}). If one want to obtain the decaying mode solution $D_{d}(a)$ from Eq. (\ref{Dask}), then one need to adopt the decaying mode initial conditions to obtain the correct coefficients  \be D_{d}(a_i) \simeq a_{i}^{-\fr{3}{2}} \hspace{0.2in} {\rm and} \hspace{0.2in} \fr{d D_{d}}{da} \Bigl|_{a_{i}} \simeq -\fr{3}{2} a_{i}^{-\fr{5}{2}} \, . \label{inid} \ee

%%%%%%%%%%%%%%%%%%%%%%%%%%%%%%%%%%%%%%%%%
\begin{center}
    \begin{table}
    \begin{tabular}{ | c | c | c | c | c | c | c | }
    \hline
    &  \multicolumn{2}{|c|}{$\Omo = 0.2$} & \multicolumn{2}{|c|}{$\Omo = 0.3$} & \multicolumn{2}{|c|}{$\Omo = 0.4$}  \\ \cline{2-7}
    $\omde $ & $c_{\sw1}$ & $c_{\sw2}$ & $c_{\sw1}$ & $c_{\sw2}$ & $c_{\sw1}$ & $c_{\sw2}$  \\ \hline
    -1/3 & 0.305329 & 0.370645 & 0.484579 & 0.546607 & 0.723051 & 0.783376  \\ \hline
    -0.8 & 0.563274 & 0.568201 & 0.704043 & 0.707640 & 0.845712 & 0.848492  \\ \hline
    -1.0 & 0.630418 & 0.631792 & 0.754267 & 0.755227 & 0.873819 & 0.874533  \\ \hline
    -1.2 & 0.680498 & 0.680868 & 0.790355 & 0.790606 & 0.893532 & 0.893715   \\ \hline
    \end{tabular}
    \caption{$c_{\sw1}$ and $c_{\sw2}$ are the values of the coefficient $c_{\sw}$ obtained from the two initial conditions of the growing mode solution, $D(a_{i}) \simeq a_{i}$ and $\fr{d D_{g}}{da} \Bigl|_{a_{i}} \simeq 1$, respectively.}
    \label{table1}
    \end{table}
\end{center}
%%%%%%%%%%%%%%%%%%%%%%%%%%%%%%%%%%%%%%%%%%%%%%%%%
Now we compare the exact growth factor in Eq. (\ref{Dask}) with the well known approximate growing mode solution \cite{Silveira}, \be D_{g}^{\sw} = c_{\sw} \Biggl( \fr{\Omo}{\Odeo} \Biggr)^{\fr{1}{3\omde}} a F \Bigl[ -\fr{1}{3 \omde}, \fr{1}{2} - \fr{1}{2\omde}, 1 - \fr{5}{6 \omde}, -\fr{\Odeo}{\Omo} a^{-3\omde} \Bigr] \, . \label{Dsw} \ee We rewrite the second term in Eq. (\ref{Dask}) using the linear transformation formula of hypergeometric function \cite{Morse}, \ba && c_{2} F \Bigl[ -\fr{1}{3\omde}, \fr{1}{2 \omde}, \fr{1}{2} + \fr{1}{6 \omde}, -\fr{\Omo}{\Odeo} a^{3\omde} \Bigr] \nonumber \\ && = c_{2} \fr{\Gamma \Bigl[\fr{1}{2}- \fr{1}{2 \omde} \Bigr] \Gamma \Bigl[1 - \fr{1}{2 \omde} \Bigr]}{\Gamma \Bigl[1 - \fr{5}{6 \omde} \Bigr]\Gamma \Bigl[\fr{1}{2}- \fr{1}{6 \omde} \Bigr]} \Biggl( \fr{\Omo}{\Odeo} \Biggr)^{\fr{1}{3\omde}} a F \Bigl[ -\fr{1}{3 \omde}, \fr{1}{2} - \fr{1}{2\omde}, 1 - \fr{5}{6 \omde}, -\fr{\Odeo}{\Omo} a^{-3\omde} \Bigr] \nonumber \\ && - \,\, c_{2} \fr{\Gamma \Bigl[-\fr{1}{2} + \fr{1}{6 \omde} \Bigr] \Gamma \Bigl[\fr{1}{2}- \fr{1}{2 \omde} \Bigr]\Gamma \Bigl[1 - \fr{1}{2 \omde} \Bigr]}{\Gamma \Bigl[- \fr{1}{3 \omde} \Bigr]\Gamma \Bigl[\fr{1}{2}- \fr{1}{3 \omde} \Bigr] \Gamma \Bigl[\fr{1}{2}- \fr{1}{6 \omde} \Bigr]}  \Biggl( \fr{\Omo}{\Odeo} \Biggr)^{\fr{3 \omde -1}{6 \omde}} a^{\fr{3 \omde -1}{2}} \nonumber \\ && \times \,\, F \Bigl[ \fr{1}{2} - \fr{1}{2\omde}, \fr{1}{2} + \fr{1}{3 \omde}, \fr{3}{2} - \fr{1}{6 \omde}, -\fr{\Omo}{\Odeo} a^{3\omde} \Bigr] \, . \label{c2F} \ea Thus, $D_{g}^{\sw}(a)$ in Eq. (\ref{Dsw}) becomes equal to $D_{g}(a)$ in Eq. (\ref{Dask}) if and only if \ba c_{1g} &=& c_{2g} \fr{\Gamma \Bigl[-\fr{1}{2} + \fr{1}{6 \omde} \Bigr] \Gamma \Bigl[\fr{1}{2}- \fr{1}{2 \omde} \Bigr]\Gamma \Bigl[1 - \fr{1}{2 \omde} \Bigr]}{\Gamma \Bigl[- \fr{1}{3 \omde} \Bigr]\Gamma \Bigl[\fr{1}{2}- \fr{1}{3 \omde} \Bigr] \Gamma \Bigl[\fr{1}{2}- \fr{1}{6 \omde} \Bigr]} \hspace{0.2in} \, \rm{and} \,\, \label{c1c2} \\  c_{\sw} &=& c_{2g} \fr{\Gamma \Bigl[\fr{1}{2}- \fr{1}{2 \omde} \Bigr] \Gamma \Bigl[1 - \fr{1}{2 \omde} \Bigr]}{\Gamma \Bigl[1 - \fr{5}{6 \omde} \Bigr]\Gamma \Bigl[\fr{1}{2}- \fr{1}{6 \omde} \Bigr]} = c_{1g} \fr{\Gamma \Bigl[- \fr{1}{3 \omde} \Bigr] \Bigl[\fr{1}{2}- \fr{1}{3 \omde} \Bigr]}{\Gamma \Bigl[-\fr{1}{2} + \fr{1}{6 \omde} \Bigr] \Gamma \Bigl[1 - \fr{5}{6 \omde} \Bigr]} \, , \label{c1csw} \ea where we use the notations that $c_{1g}$ and $c_{2g}$ are the values of coefficients $c_{1}$ and $c_{2}$ obtained from the growing mode initial conditions in Eq. (\ref{inig}).
%%%%%%%%%%%
%\begin{center}
%\begin{figure}
%\vspace{1.5cm}
%\centerline{
%\psfig{file=c1oc2.eps, width=6.5cm} \psfig{file=csw.eps, width=6.5cm} }
%\vspace{-0.5cm}
%\caption{ The relationship between the coefficients given in Eqs. (\ref{c1c2}) and (\ref{c1csw}). a) The ratio of $c_{1g}$ to $c_{2g}$. b) The ratios of $c_{\sw}$ to $c_{1g}$ and $c_{2g}$ (from top to bottom).} \label{fig2}
%\end{figure}
%\end{center}
%%%%%%%%%%%%%%
$D_{\sw}(a)$ given in Eq. (\ref{Dsw}) have several problems. First, $D_{\sw}(a)$ contains only one integral constant $c_{\sw}$ even though it is obtained from the second order differential equation. This problem might be solved if the two integral constants in the general solution (\ref{Dask}) satisfy the conditions in Eqs. (\ref{c1c2}) and (\ref{c1csw}) simultaneously. In other word, $D_{\sw}(a)$ would be an exact growing mode solution if the values of $c_{sw}$ obtained from both initial conditions (\ref{inig}) are the same. We denote that $c_{\sw1}$ and $c_{\sw2}$ are the values of $c_{\sw}$ obtained from the growing mode initial conditions $D_{g}(a_i) \simeq a_{i}$ and $\fr{d D_{g}}{da} \Bigl|_{a_{i}} \simeq 1$, respectively. As we show in Tab. \ref{table1}, $c_{\sw1}$ and $c_{\sw2}$ show discrepancies for the different models. As $\omde$ decreases, the difference between the two coefficients also decreases. The same effects happen when $\Omo$ is big. Thus, $D_{\sw}$ is a good approximate solution for the small value of $\omde$ and the big value of $\Omo$.
%%%%%%%%%%%%%%%%%%%%%%%%%%%%%%%%%%%%%%%%%%%%%%%%%
\begin{center}
    \begin{table}
    \begin{tabular}{ | c | c | c | c | c | c | c | c | c |}
    \hline
    $\omde$ & $\xi$ & $c_{\sw1}$ & $c_{\sw2}$ & $D_{\sw1}(1)$ & $D_{\sw2}(1)$ & $D_{g}(1)$ & $f_{\sw}(a_{i})$ & $f(a_{i})$ \\ \hline
    -0.4 & -0.0486 & 0.589752 & 0.606695 & 0.624667 & 0.642613 & 0.631619 & 0.924788 & 0.951356\\ \hline
    -0.8 & -0.0011 & 0.705845 & 0.708663 & 0.733052 & 0.735979 & 0.734198 & 0.994917 & 0.998890\\ \hline
    -1.0 & 0.0000 & 0.754267 & 0.755227 & 0.779311 & 0.780303 & 0.779699 & 0.998730 & 1.000000 \\ \hline
\end{tabular}
    \caption{$\xi$, $c_{\sw1}$, $c_{\sw2}$, $D_{\sw1}(a=1)$, and $D_{\sw2}(a=1)$ for the different values of $\omde$ when we choose $\Omo = 0.3$ and $a_{i} = 0.1$. $D_{\sw1}(1)$ and $D_{\sw2}(1)$ are the present values of the growth factors when we choose $c_{\sw1}$ and $c_{\sw2}$, respectively. $D_{g}(1)$ is the present value of the exact growth factor. $f_{\sw}(a_{i})$ and $f(a_{i})$ correspond to the initial values of the growth index obtained from $D_{\sw}$ and $D_{g}$, respectively.}
    \label{table2}
    \end{table}
\end{center}
%%%%%%%%%%%%%%%%%%%%%%%%%%%%%%%%%%%%%%%%%%%%%%%%%

\begin{center}
\begin{figure}
\vspace{1.5cm}
\centerline{
\psfig{file=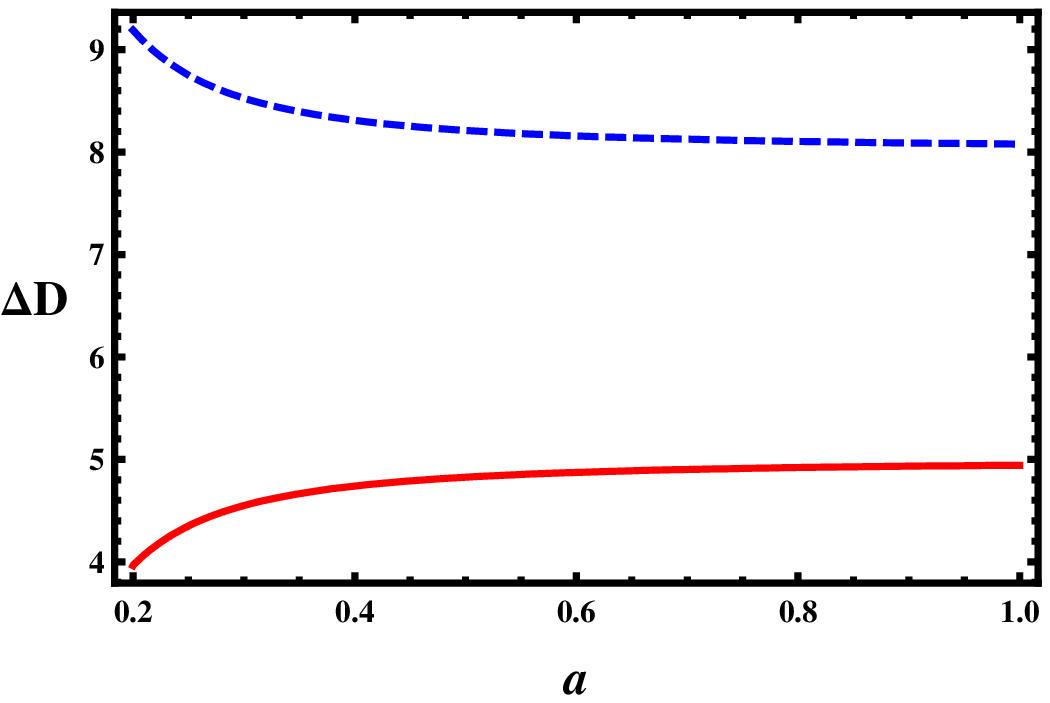, width=6.5cm} \psfig{file=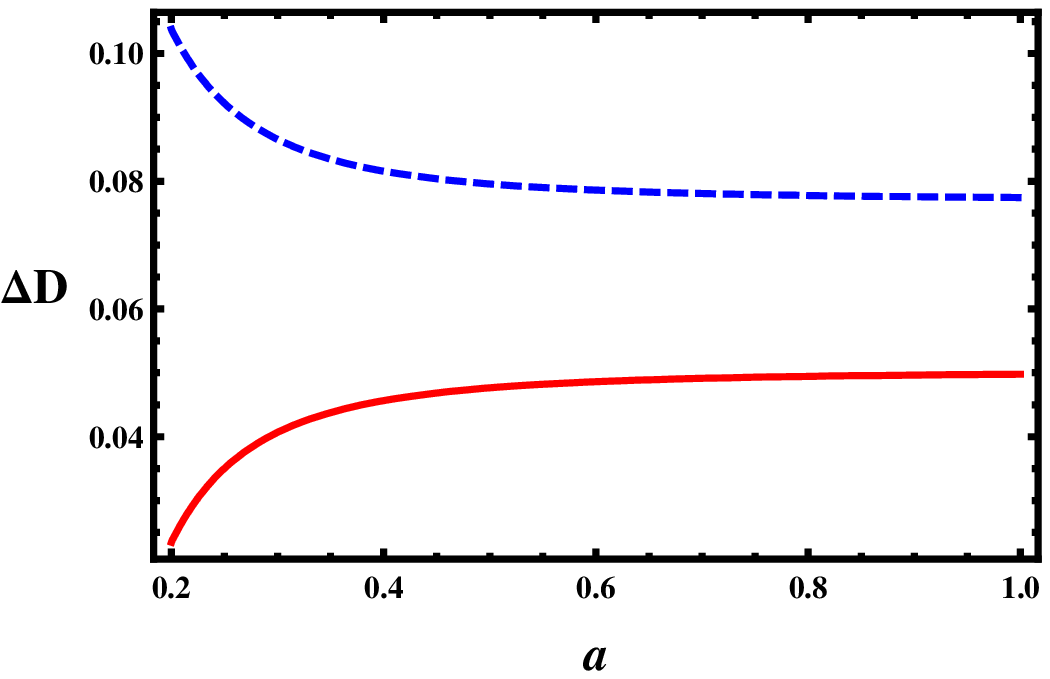, width=6.5cm} }
\vspace{-0.5cm}
\caption{ Errors of $D_{\sw1}$ and $D_{\sw2}$. a) $|D_{g} - D_{\sw2}| / D_{g} \times 100$ (\%) and  $|D_{g} - D_{\sw1}| / D_{g} \times 100$ (\%) (from top to bottom) for $\omde = -0.4$ and $\Omo = 0.2$. b) $|D_{g} - D_{\sw2}| / D_{g} \times 100$ (\%) and  $|D_{g} - D_{\sw1}| / D_{g} \times 100$ (\%) (from top to bottom) for $\omde = -1.0$ and $\Omo = 0.3$.} \label{fig1}
\end{figure}
\end{center}

One may suspect that this discrepancy between $c_{\sw1}$ and $c_{\sw2}$ might be due to the choices of initial conditions. We investigate this as follows. The exact values of initial conditions can be obtained numerically from Eq. (\ref{daD}), \be D_{g}(a_i) = a_{i}^{1 + \xi} \hspace{0.2in} {\rm and} \hspace{0.2in} \fr{d D_{g}}{da} \Bigl|_{a_{i}} = (1 + \xi) a_{i}^{\xi} \, , \label{inigexact} \ee where $\xi$ indicates the deviation of the growth factor from the linear growth $a_i$ at the initial epoch. We show the magnitudes of these $\xi$s for the different DE models in Tab. \ref{table2}. In this table, we choose $\Omo = 0.3$ and $a_{i} = 0.1$. As $\omde$ decreases, the value of $\xi$ also decreases because there is more matter component at the initial epoch $a_{i}$ for the smaller values of $\omde$. Thus, $D_{g}(a_{i})$ is close to $a_{i}$ for the smaller value of $\omde$. We also shows the values of $c_{\sw1}$ and $c_{\sw2}$ for different DE models with initial conditions in Eq. (\ref{inigexact}). If we compare $c_{\sw1}$ and $c_{\sw2}$ values in Tab. \ref{table1} with those in Tab. \ref{table2}, then we find that the discrepancies in the two values are not removed even with the exact values of initial conditions. Thus, the deviations in $c_{\sw1}$ and $c_{\sw2}$ are the intrinsic problem of the solution $D_{\sw}$ and irrelevant to the accuracies of initial conditions. We define $D_{\sw1}$ and $D_{\sw2}$ as the growth factor obtained from Eq. (\ref{Dsw}) when we choose the coefficient $c_{\sw}$ as $c_{\sw1}$ and $c_{\sw2}$, respectively. The discrepancies in the present values of the growth factor $D_{\sw1}$ and $D_{\sw2}$ are decreased as $\omde$ is decreased.  $D_{\sw}(a)$ is a good approximate solution for the small values of $\omde$ and big values of $\Omo$. We show this in Fig. \ref{fig1}. When $\omde$ is bigger than $-1$ and $\Omo$ is small, $D_{\sw1}$ and $D_{\sw2}$ show the big discrepancies with the correct $D_{g}$ as shown in the left panel of Fig. \ref{fig1}. The dashed, solid lines correspond to percentage errors of $(D_{g} - D_{\sw1})/D_{g}$ and $(D_{g} - D_{\sw2})/D_{g}$, respectively when $\omde = -0.4$ and $\Omo = 0.2$. Thus, the errors at $a \geq 0.6$ are as big as $8$ \% and $5$ \% in each case. However, when $\omde$ is close and smaller than $-1$ and $\Omo$ is big, both $D_{\sw1}$ and $D_{\sw2}$ are very close to $D_{g}$ as shown in the right panel of Fig. \ref{fig1} where we use $\omde = -1.0$ and $\Omo = 0.3$. The lines indicate same percentage errors as in the left panel of Fig. \ref{fig1}. As shown in the figure, the errors are sub percentages in this case.
%%%%%%%%%%%
%%%%%%%%%%%%%%

Second, one is not able to separate the growing mode solution and the decaying mode solution in $D(a)$. So far, there are only two known possible cases for separating the two modes when $\omde = -\fr{1}{3}$ and $-1$ \cite{Dodelson}. This separation is impossible for general values of $\omde$ as shown above. Thus, $D_{\sw}$ is not the correct growing mode solution. $D(a)$ itself in Eq. (\ref{Dask}) is the solution of the equation (\ref{daD}), and this solution cannot be separated as the growing mode solution or the decaying one. As we explained in the above, after we obtain the general solution of the equation (\ref{daD}), the solution $D(a)$ can be interpreted as the growing or the decaying mode solution by applying the appropriate initial conditions given in Eqs. (\ref{inig}) and (\ref{inid}) to the general solution in Eq. (\ref{Dask}).

The third problem is related to the growth index $f = \fr{d \ln D}{d \ln a}$. If we choose  $D_{\sw}$ as the growing mode solution, then the growth index becomes \be f_{\sw} = \fr{d \ln D_{\sw}}{d \ln a} =  \fr {\ln \Biggl[ a F \Bigl[ -\fr{1}{3 \omde}, \fr{1}{2} - \fr{1}{2\omde}, 1 - \fr{5}{6 \omde}, -\fr{\Odeo}{\Omo} a^{-3\omde} \Bigr] \Biggr]}{d \ln a} \, . \label{fsw} \ee Since $D_{\sw}$ has only one coefficient, the growth index obtained from $D_{\sw}$ is independent of $c_{\sw}$. If we choose the exact values of initial conditions given in Eq. (\ref{inigexact}), then the value of the growth index at the initial epoch will become \be f(a=a_{i}) = 1 + \xi + {\cal O} (\xi^3) + \cdots \, . \label{fai} \ee Therefore, $f_{\sw}(a=a_{i})$ is not same as $f(a_{i})$ given in Eq. (\ref{fai}). This problem also happens when we choose the approximate initial conditions (\ref{inig}). Thus, the value of the growth index obtained from $D_{\sw}$ shows the intrinsic discrepancies with that obtained from the correct growth factor $D_{g}$ as shown in Tab. \ref{table2}. We find that the present value of $f$ for $\Omo = 0.4$ should be close to $0.6$ independent of $\omde$ and thus Fig. $3$ in Ref. \cite{Silveira} is incorrect.

Recently, we have also obtained the exact analytic solution of $D(a)$ for $\omde = -1$ \cite{SK1}. There we have found that the solution of $D_{g}$ for $\omde = -1$ is given by \be D_{g}^{L}(a) = c_{1}^{L} Q^{\fr{2}{3}} a^{-2} F \Bigl[1, \fr{1}{6}, \fr{5}{3}, - Q a^{-3} \Bigr] + c_{2}^{L} \sqrt{1 + Q a^{-3}} \, . \label{DgL} \ee The form of $D_{g}^{L}(a)$ looks quite different from $D_{g}(a)$ in Eq. (\ref{Dask}). However, when $\omde = -1$, the general solution $D(a)$ becomes \ba D(a) |_{\omde = -1} &=& c_{1} \Biggl( \fr{\Omo}{\Odeo} \Biggr)^{\fr{2}{3}} a^{-2} F \Bigl[ 1, \fr{1}{6}, \fr{5}{3}, -\fr{\Omo}{\Odeo} a^{-3} \Bigr] +  c_{2} F \Bigl[ \fr{1}{3}, - \fr{1}{2}, \fr{1}{3}, -\fr{\Omo}{\Odeo} a^{-3} \Bigr] \nonumber \\ &=& c_{1} \Biggl( \fr{\Omo}{\Odeo} \Biggr)^{\fr{2}{3}} a^{-2} F \Bigl[ 1, \fr{1}{6}, \fr{5}{3}, -\fr{\Omo}{\Odeo} a^{-3} \Bigr] +  c_{2} F \Bigl[ - \fr{1}{2}, \fr{1}{3}, \fr{1}{3}, -\fr{\Omo}{\Odeo} a^{-3} \Bigr] \nonumber \\ &=& c_{1} \Biggl( \fr{\Omo}{\Odeo} \Biggr)^{\fr{2}{3}} a^{-2} F \Bigl[ 1, \fr{1}{6}, \fr{5}{3}, -\fr{\Omo}{\Odeo} a^{-3} \Bigr] +  c_{2} \sqrt{1 + \fr{\Omo}{\Odeo} a^{-3}} = D_{g}^{L}(a) \, , \label{Daw1} \ea where we use the relation $F \Bigl[j, k, j, -Y \Bigr] = F \Bigl[ k, j, j, -Y \Bigr] = \sqrt{1 + Y}$ in the second and the third equalities \cite{Morse}. Thus, the solution $D_{g}^{L}(a)$ given in Eq. (\ref{DgL}) is one of the particular solutions of $D(a)$ when $\omde = -1$. We are also able to obtain the particular solution of $D(a)$ when $\omde = -\fr{1}{3}$ by using the same relation. \ba D(a) |_{\omde = -\fr{1}{3}} &=& c_{1} \Biggl( \fr{\Omo}{\Odeo} \Biggr) a^{-1} F \Bigl[ 2, -\fr{1}{2}, 2, -\fr{\Omo}{\Odeo} a^{-1} \Bigr] +  c_{2} F \Bigl[ 1, - \fr{3}{2}, 0, -\fr{\Omo}{\Odeo} a^{-1} \Bigr] \nonumber \\ &=& c_{1} \Biggl( \fr{\Omo}{\Odeo} \Biggr) a^{-1} \sqrt{1 + \fr{\Omo}{\Odeo} a^{-1}} + c_{2} F \Bigl[ 1, - \fr{3}{2}, 0, -\fr{\Omo}{\Odeo} a^{-1} \Bigr] \nonumber \\ &=& c_{1} \Biggl( \fr{\Omo}{\Odeo} \Biggr) a^{-1} \sqrt{1 + \fr{\Omo}{\Odeo} a^{-1}} \nonumber \\ &+& c_{2} \Biggl( -1 - 3 \fr{\Omo}{\Odeo} a^{-1} + 3 \fr{\Omo}{\Odeo} a^{-1} \sqrt{1 + \fr{\Omo}{\Odeo} a^{-1}} \arctanh \Bigl[ \sqrt{\fr{\Omo}{\Odeo} a^{-1}} \Bigr] \Biggr) \, , \label{Daw13} \ea where $\arctanh$ is the inverse hyperbolic tangent function.

%%%%%%%%%%%%%%%%%%%%%%%%%%%%%%%%%%%%%%%%%%%%%%%%%%%%%%%%%%%%%%%%%%%%%%%%%
\section{Comparison with known approximate solutions}
\setcounter{equation}{0}
%%%%%%%%%%%%%%%%%%%%%%%%%%%%%%%%%%%%%%%%%%%%%%%%%%%%%%%%%%%%%%%%%%%%%%%%%
There are several well known approximate analytic forms of the growth factor \cite{0508156,Carroll,0303112}. For the cosmological constant ({\it i.e.} $\omde = -1$), the well known approximate form of the growth factor at present is given by \cite{Carroll} \be D_{\cpt}^{0} = \fr{5 \Omo}{2} \Biggl[ \Bigl( \Omo \Bigr)^{\fr{4}{7}} - \OLo + \Bigl( 1 + \fr{\Omo}{2} \Bigr) \Bigl( 1 + \fr{\OLo}{70} \Bigr) \Biggr]^{-1} \, , \label{Dcpt0} \ee
where $\OLo$ is the present value of the energy density contrast ($\rho_{\Lambda}^{0} / \rho_{\rm{cr}}^{0}$) of the cosmological constant $\Lambda$. One is not able to obtain the growth index from the $D_{\cpt}^{0}$ because it is a constant. Thus, the approximate analytic form of the growth index $f_{\lahav}$ is given separately in Ref. \cite{Lahav} : \be f_{\lahav}(a) = \Biggl[\fr{\Oma}{\Oma + \OLa} \Biggr]^{\fr{4}{7}} \, . \label{flahav} \ee The above solution is generalized to any value of $a$ in Ref. \cite{0508156} : \be D_{\cpt}(a) = \fr{5 \Oma}{2}  a \Biggl[ \Bigl( \Oma \Bigr)^{\fr{4}{7}} - \OLa + \Bigl( 1 + \fr{\Oma}{2} \Bigr) \Bigl( 1 + \fr{\OLa}{70} \Bigr) \Biggr]^{-1} \, . \label{Dcpt} \ee We compare this solution $D_{\cpt}$ with the exact one in Eq. (\ref{Daw1}).
%%%%%%%%%%%
\begin{center}
\begin{figure}
\vspace{1.5cm}
\centerline{
\psfig{file=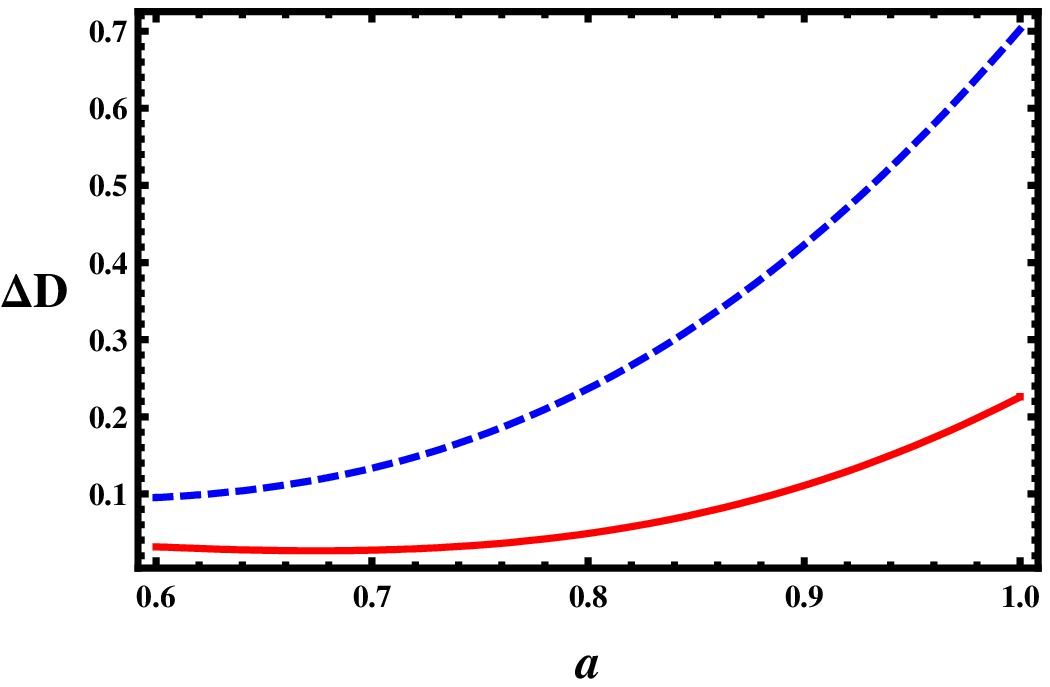, width=6.5cm} \psfig{file=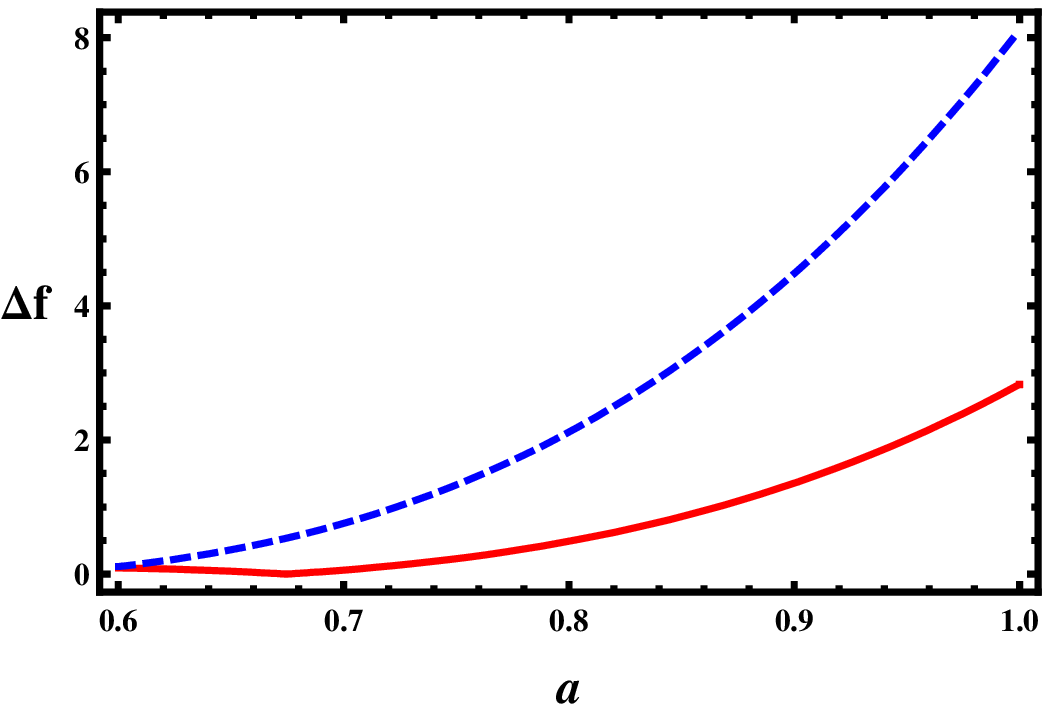, width=6.5cm} }
\vspace{-0.5cm}
\caption{ a) Relative errors of $|D_{g} - D_{\cpt}| / D_{g} \times 100$ (\%) when $\Omo = 0.2$ and $0.3$ (from top to bottom). b) Relative errors of $|f - f_{\cpt}| / f \times 100$ (\%) when $\Omo = 0.2$ and $0.3$ (from top to bottom).} \label{fig2}
\end{figure}
\end{center}
%%%%%%%%%%%%%%

We show the relative errors of $D_{\cpt}$ and $f_{\cpt}$ compared to the exact solutions $D_{g}$ and $f$ when $\Omo = 0.2$ and $0.3$ in Fig. \ref{fig2}. The errors of the analytic approximate solution $D_{\cpt}$ are smaller than $1$ \% as shown in the left panel of Fig. \ref{fig2}. The dashed and solid lines are $|D_{g} - D_{\cpt}| / D_{g} \times 100$ (\%) when $\Omo = 0.2$ and $0.3$, respectively. $f_{\cpt}(a)$ is also obtained from $D_{\cpt}$ in Eq. (\ref{Dcpt}). We compare the evolution of $f_{\cpt}$ with that of $f$ in the right panel of Fig. \ref{fig2}. The solid line describes the correct $f$ obtained from $D_{g}$. The dotted line shows the evolution $f_{\cpt}$ obtained from $D_{\cpt}$. The error of $f_{\cpt}$ at present is about $4$ \% only when we use $\Omo = 0.3$.

There is also another approximate analytic solution $D_{\bp}$ for the general values of $\omde$ \cite{0508156,0303112}. This solution is obtained from the well known parametrization of the growth index and its parameter in Ref. \cite{WS}, \ba f &=& \fr{d \ln D_{g}}{d \ln a} = \Oma^{\gamma_{\ws}} \, , \label{fWS} \\ {\rm where} \,\,\, \gamma_{\ws} &\simeq& \fr{3 (1 - \omde)}{5 - 6 \omde } + \fr{3}{125} \fr{(1-\omde)(1-\fr{3\omde}{2})}{(1-\fr{6\omde}{5})^3} ( 1 - \Oma) \, . \label{gammaWS} \ea The approximate growth factor $D_{\bp}$ is known as the extension of $D_{\cpt}$ in Eq. (\ref{Dcpt}) for the general $\omde$ and is given by \cite{0303112} \be D_{\bp}^{0}(a) = \fr{5\Omo}{2}  a \Biggl[ \Bigl( \Omo \Bigr)^{\gamma_{\ws}^{0}} - \Odeo + \Bigl( 1 + \fr{\Omo}{2} \Bigr) \Bigl( 1 + {\cal A} \Odeo \Bigr) \Biggr]^{-1} \, , \label{Dbp0} \ee where $\gamma_{\ws}^{0}$ is the approximate form of the growth index parameter by choosing $\Oma = \Omo$ in Eq. (\ref{gammaWS}) and ${\cal A}$ is the fitting coefficient in Ref. \cite{0508156,0303112} \ba \gamma_{\ws}^{0} &\simeq& \fr{3 (1 - \omde)}{5 - 6 \omde} + \fr{3}{125} \fr{(1-\omde)(1-\fr{3\omde}{2})}{(1-\fr{6\omde}{5})^3} ( 1 - \Omo) \, , \label{gammaWS0} \\ {\cal A} &\simeq& 1.742 + 3.343 \omde + 1.615 \omde^2 \,  \hspace{0.2in} {\rm when} \,\, \omde \geq -1 \label{calAB} \\ &\simeq& -\fr{0.28}{\omde + 0.08} - 0.3 \,  \hspace{0.8in} {\rm when} \,\, \omde < -1 \, . \label{calAP} \ea
%%%%%%%%%%%
\begin{center}
\begin{figure}
\vspace{1.5cm}
\centerline{
\psfig{file=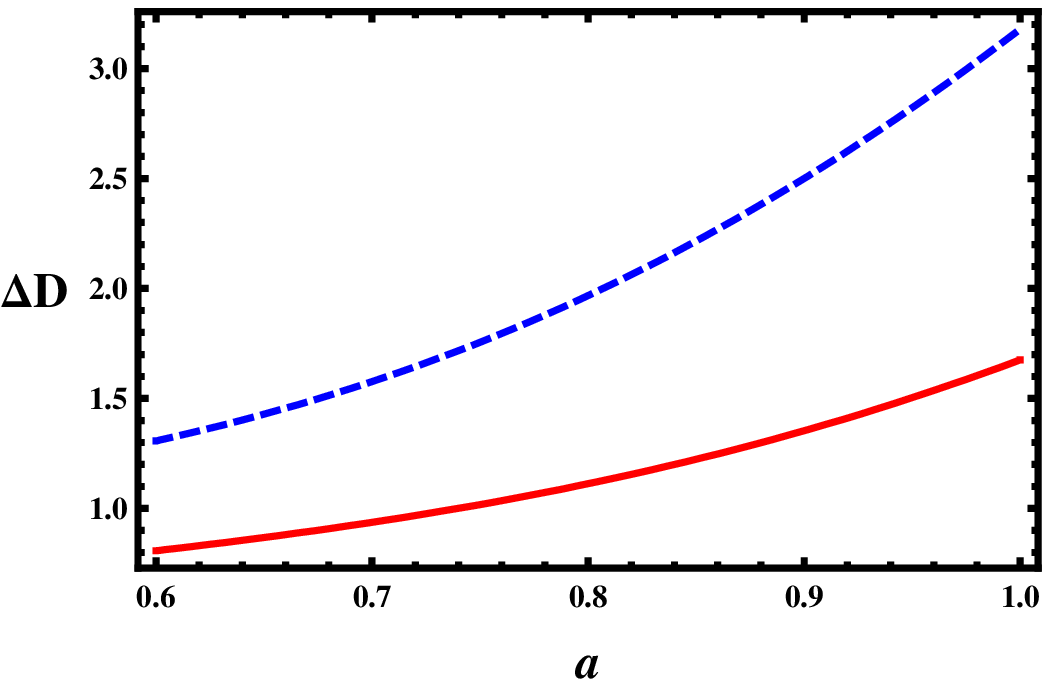, width=6.5cm} \psfig{file=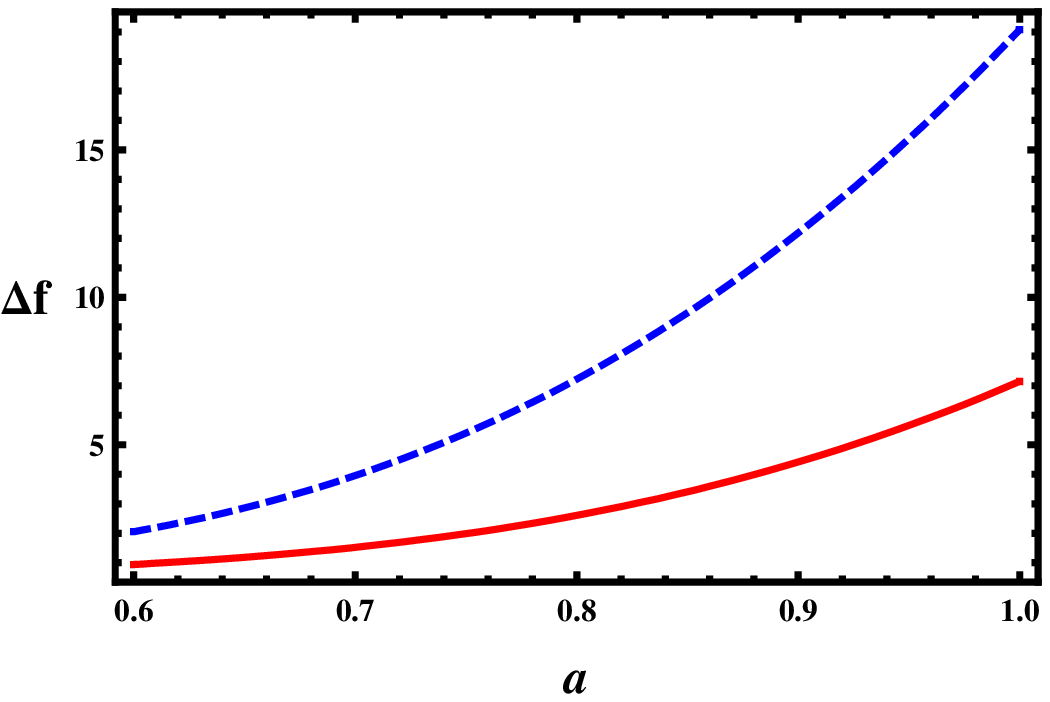, width=6.5cm} }
\vspace{-0.5cm}
\caption{ a) Relative errors of $|D_{g} - D_{\bp}| / D_{g} \times 100$ (\%) when $\Omo = 0.2$ and $0.3$ (dashed and solid lines, respectively) for $\omde = -0.8$. b) Relative errors of $|f - f_{\bp}| / f \times 100$ (\%) (with same notation) for same $\omde$ and $\Omo$s.} \label{fig3}
\end{figure}
\end{center}
%%%%%%%%%%%%%%

Again the above solution $D_{\bp}^{0}$ is generalized as \cite{0508156} \be D_{\bp}(a) = \fr{5\Oma}{2}  a \Biggl[ \Bigl( \Oma \Bigr)^{\gamma_{\ws}} - \Odea + \Bigl( 1 + \fr{\Oma}{2} \Bigr) \Bigl( 1 + {\cal A} \Odea \Bigr) \Biggr]^{-1} \, . \label{Dbp} \ee Even though the value of $D_{\bp}$ at any epoch $a$ is very close to that of $D_{g}$, its evolution behavior is quite different from that of $D_{g}$. We show this in Fig. \ref{fig3}. In the left panel of Fig. \ref{fig3}, the dashed and solid lines correspond to $|D_{g} - D_{\bp}| / D_{g} \times 100$ \% where $D_{\bp}$ with ${\cal A}$ given in Eq. (\ref{calAB}) when $\Omo = 0.2$ and $0.3$ for $\omde = -0.8$. Even though the error in $D_{\bp}$ at present is about $3$ \%, there is the discrepancy in cosmological evolution behaviors of $D_{g}$ and $D_{\bp}$. This discrepancy is clear when we compare the growth index $f$ and $f_{\bp}$ as shown in the right panel of Fig. \ref{fig3}. Again the dashed and solid lines describe $|f - f_{\bp}| / f \times 100$ \% for the same values of $\Omo$s and $\omde$ as in the left panel of Fig. \ref{fig3}. The error in $f_{\bp}$ is more than $20$ \% at present.

As we show in Figs. \ref{fig2} and \ref{fig3}, one might be able to obtain the value of the growth factor with small error from the approximate analytic solutions of the growth factor. However, one needs to pay attention when one considers the growth index. Especially, $D_{\bp}$ might not be used to compare with observations because of the incorrect behavior of $f_{\bp}$ obtained from the approximate analytic solution $D_{\bp}$ for some DE models.

\begin{center}
\begin{figure}
\vspace{1.5cm}
\centerline{
\psfig{file=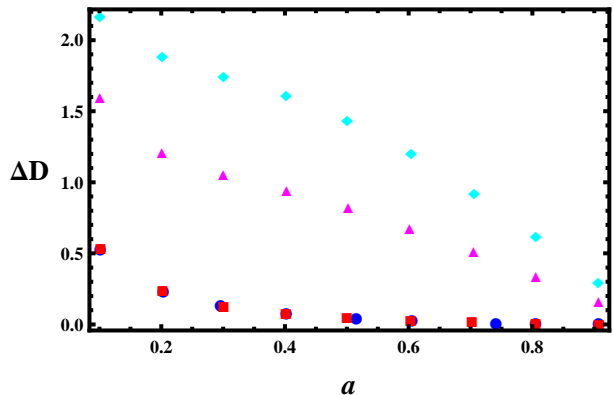, width=8cm}}
\vspace{-0.5cm}
\caption{ The relative errors of the analytic sub-horizon growth factor compared to the numerical one obtained from the CMBFAST. The square, circle, triangle, and diamond dots represent the relative errors when $\omde = -0.9, -1.0, -1.1$, and $-1.2$, respectively.} \label{fig4}
\end{figure}
\end{center}

We also compare the sub-horizon growth factor obtained from the Eq. (\ref{daD}) with the exact numerical one obtained from the numerical code, CMBFAST \cite{CMBFAST}. As we show in the Fig. \ref{fig4}, the relative errors of analytic sub-horizon growth factor is less than $1.5$ \% when we consider $a \geq 0.6$ in almost all interest cases. The square, circle, triangle, and diamond dots represent the relative error of analytic solution in Eq. (\ref{Dask}) compared to the one obtained from CMBFAST when $\omde = -0.9, -1.0, -1.1$, and $-1.2$, respectively. We use the numerical solution at wavenumber $k = 0.09 h Mpc{-1}$. The result does not change for the other sub-horizon scale numerical solutions.
%%%%%%%%%%%
%%%%%%%%%%%%%%
We obtain the analytic solutions for $D_{g}$ and $f$ that are exact for any DE model. And these solutions give the exact theoretical values of observable quantities. However, the exact solutions are limited to the constant values of $\omde$. Thus, we need to investigate the generalization of the solutions to more general cases including time-varying $\omde$. We will explain the possible extensions of them in the following section.

%%%%%%%%%%%%%%%%%%%%%%%%%%%%%%%%%%%%%%%%%%%%%%%%%%%%%%%%%%%%%%%%%%%%%%%%%
\section{Applications for $D_{g}(a)$ and $f(a)$ to time-varying $\omde$}
\setcounter{equation}{0}
%%%%%%%%%%%%%%%%%%%%%%%%%%%%%%%%%%%%%%%%%%%%%%%%%%%%%%%%%%%%%%%%%%%%%%%%%
It is well known that the time-dependence of $\omde$ is extremely difficult to
discern because the dark energy is dynamically unimportant at the redshifts
where $\omde$ departs from its low z value. In addition, for the substantial
changes in $\omde$ at low redshift, there is always a constant $\omde$ that produces
very similar evolution of all of the observables simultaneously \cite{0112221, 0112526}. Also
this analytic solution can provide useful templates to study the structure
growth in dark energy models with time varying equation of state. The analytic solution is also very useful when one calculates the abundance of galaxy clusters as a function of redshift.

\begin{center}
\begin{figure}
\vspace{1.5cm}
\centerline{
\psfig{file=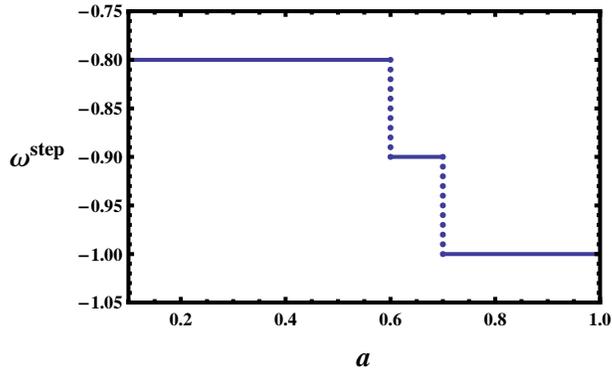, width=8cm}}
\vspace{-0.5cm}
\caption{ A specific model of Eq. (\ref{wstep}).} \label{fig5}
\end{figure}
\end{center}

\begin{center}
\begin{figure}
\vspace{1.5cm}
\centerline{
\psfig{file=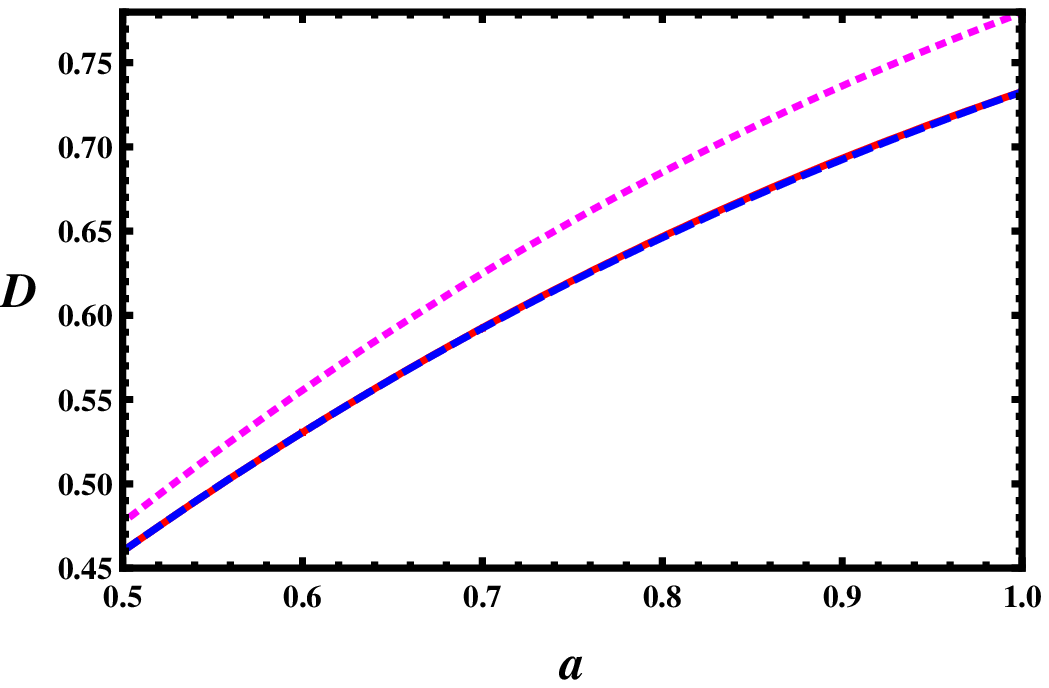, width=6.5cm} \psfig{file=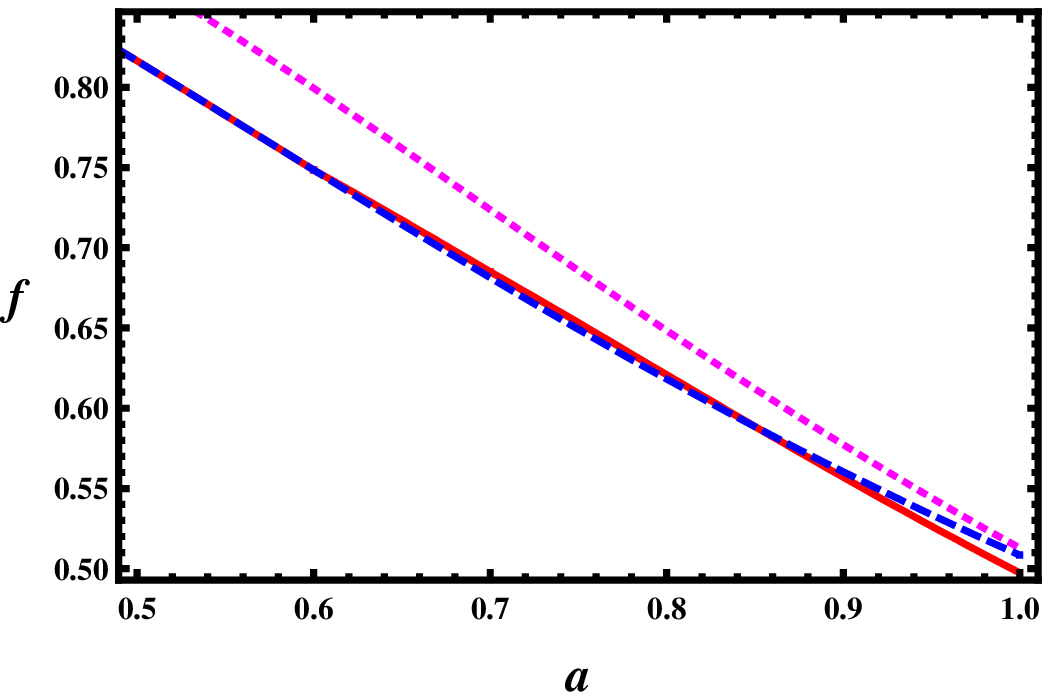, width=6.5cm} }
\vspace{-0.5cm}
\caption{ a) Cosmological evolutions of $D_{g}(a)$ for $\omde = -1.0$, $\omde^{\step}$, and $-0.8$ (from top to bottom) when $\Omo = 0.3$. b) Evolutions of $f(a)$ for $\omde = -1.0$, $-0.8$, and $\omde^{\step}$ (from top to bottom) for $\Omo = 0.3$.} \label{fig6}
\end{figure}
\end{center}

Even though the growth factor obtained in Eq. (\ref{Dask}) is only true for the constant $\omde$, we are able to apply this solution to the time-varying $\omde$ by interpolating between models with constant $\omde$ \cite{0209093,0508156,08040413,08073140}. For this purpose, we choose the sum of the step functions $\theta$ of $\omde(a)$ to probe the evolutions of $D_{g}(a)$ and $f(a)$, \be \omde^{\step}(a) = \sum_{j} \omde(j) \theta(a - a_{j}) \, , \label{wstep} \ee where $\omde(j)$ is the arbitrary value we need to fit from the background evolution observations. We use a specific model of this, $\omde^{\step} = -0.8 \theta (a) - 0.1 \theta(a - 0.6) -0.1 \theta (a - 0.7)$, for the demonstration as shown in Fig. \ref{fig5}. The values of $\omde(j)$ and $a_{j}$ are related to the values of the $\omde$ parameterization which produce the proper background evolution like $H(a)$ \cite{08040413}. Also, one is able to extend this parametrization to more general cases by putting more steps and/or different values of $\omde(j)$.
%%%%%%%%%%%%%%%%%%%%%%%%%%%%%%%%%%%%%%%%%%%%%%
%%%%%%%%%%%%%%
The advantages of this parametrization of $\omde$ are the followings. Even though the EOS is a discontinuous function of $a$ ({\it i.e.} $z$), the physical quantities like $H(a)$, $D_{g}(a)$, and $f(a)$ obtained from this $\omde^{\step}$ are smooth functions \cite{0209093}. We are able to obtain the smooth functions $D_g(a)$ and $f(a)$ by solving for the proper values of $c_{1}$ and $c_{2}$ in Eq. (\ref{Dask}) at each interval. This is shown in Tab. \ref{table3}. Also the observations constrain the physical quantities in the specific interval of $a$. Thus, the parameterization of $\omde$ in Eq. (\ref{wstep}) is a good one to probe the properties of $\omde$ when compared to the observations.
%%%%%%%%%%%%%%%%%%%%%%%%%%%%%%%%%%%%%%%%%
\begin{center}
    \begin{table}
    \begin{tabular}{ | c | c | c | c | c | c | }
    \hline
    \multicolumn{2}{|c|}{$0.1 \leq a \leq 0.6$} & \multicolumn{2}{|c|}{$0.6 \leq a \leq 0.7$} & \multicolumn{2}{|c|}{$0.7 \leq a \leq 1.0$}  \\ \hline
    $D_g(0.1)$ & $\fr{d D_g}{da}|_{0.1}$ & $D_g(0.6)$ & $\fr{d D_g}{da}|_{0.6}$ & $D_g(0.7)$ & $\fr{d D_g}{da}|_{0.7}$ \\ \hline
    0.1 & 1 & 0.530531 & 0.661798 & 0.592623 & 0.579974 \\ \hline
    $c_{1}$ & $c_{2}$ & $c_{1}$ & $c_{2}$ & $c_{1}$ & $c_{2}$  \\ \hline
    1.09339 & -1.39169 & 1.04914 & -1.07846 & 1.01286 & -0.872337  \\ \hline
    \end{tabular}
    \caption{$c_{1}$ and $c_{2}$ are the values of the coefficients obtained from $D(a = a_j)$ and $\fr{d D(a)}{da} \Bigl|_{a= a_{j}}$ at each interval.}
    \label{table3}
    \end{table}
\end{center}
%%%%%%%%%%%%%%%%%%%%%%%%%%%%%%%%%%%%%%%%%%%%%%%%%
We show the cosmological evolutions of $D_{g}$ and $f$ in Fig. \ref{fig6} for the different DE models. The evolutions of the growth factors $D_{g}$ for $\omde = -1.0$, $\omde^{\step}$, and $-0.8$ are described as dotted, solid, and dashed lines, respectively in the left panel of Fig. \ref{fig6}. The present values of $D_g$ are $(0.7797, 0.7328, 0.7327)$ for $\omde = -1.0$, $\omde^{\step}$, and $-0.8$, respectively. The evolutions of the $D_{g}$ for $\omde^{\step}$ and $\omde = -0.8$ are quite similar to each other because of the specific choices for values of $\omde^{\step}$ in Eq. (\ref{wstep}). If we have the shorter period of $a_{j =1}$ for $\omde = -0.8$ in $\omde^{\step}$, then we may have the different evolution of $D_{g}$ from the different choices of $\omde^{\step}$. Also the cosmological evolutions of $f$ are depicted as the dotted, solid, and dashed lines for $\omde = -1.0$, $-0.8$, and $\omde^{\step}$, respectively in the right panel of Fig. \ref{fig6}. We obtain the present values of $f$, $(0.5128, 0.5084, 0.4974)$ when $\omde = -1.0$, $-0.8$, and $\omde^{\step}$, respectively. Thus, we obtain very interesting features of $D_{g}$ and $f$ from these DE models. Even though the present values of $\omde = -1.0$ and $\omde^{\step}$ are equal, the evolution behaviors of $D_{g}$ and $f$ are quite different for these two models as shown in Fig. \ref{fig6}. $D(a=1)$ values are different by as large as $6$ \% and the difference in $f(a=1)$ is about $3$ \%. Thus, we may have a good chance to tell whether $\omde$ is a constant or not by investigating $D_{g}(a)$ and $f(a)$ at different $a$ intervals.
%%%%%%%%%%%
%%%%%%%%%%%%%%

%%%%%%%%%%%%%%%%%%%%%%%%%%%%%%%%%%%%%%%%%%%%%%%%%%%%%%%%%%%%%%%%%%%%%%%%%
\section{Conclusion and discussion}
\setcounter{equation}{0}
%%%%%%%%%%%%%%%%%%%%%%%%%%%%%%%%%%%%%%%%%%%%%%%%%%%%%%%%%%%%%%%%%%%%%%%%%
We have analyzed the properties of the exact analytic solution of sub-horizon scale matter density perturbation ({\it i.e.} growth factor) for the general dark energy models with its equation of state $\omde$ being constant. From the comparison of this solution $D_{g}$ with the well known approximate analytic solution $D_{g}^{\sw}$, we have found that $D_{g}^{\sw}$ is a good approximate solution of $D_{g}$ for the concordance model. $D_{g}$ can be expressed with the slightly different functional forms for the specific values of $\omde$. Especially, we have explicitly shown the alternative forms of $D_{g}$ when $\omde = -1$ and $-\fr{1}{3}$. The two solutions in Refs. \cite{SK1} and \cite{SK3} are equivalent when $\omde = -1$ or $-\fr{1}{3}$ even though they look quite different.

We have scrutinized the several well known approximate analytic forms of the growth factor. $D_{\cpt}$ is the one with the dark energy being the cosmological constant and $D_{\bp}$ is the extension of $D_{\cpt}$ for the general dark energy models with constant $\omde$. $f_{\cpt}$ and $f_{\bp}$ are the growth indices obtained from $D_{\cpt}$ and $D_{\bp}$, respectively. When the dark energy is the cosmological constant, $D_{\cpt}$ and $f_{\cpt}$ are very close to the correct $D_{g}$ and $f$. However, $D_{\bp}$ and $f_{\bp}$ show the discrepancies with the correct $D_{g}$ and $f$ for some dark energy models. Especially, the error in $f_{\bp}$ for $\omde = -0.8$ and $\Omo = 0.3$ is as large as $20$\% at present.

The approximate analytic solution $D_{\sw}$, the approximate analytic forms $D_{\cpt}$ and $D_{\bp}$ are good approximate solutions of the exact $D_{g}$ for the concordance model. However, all of them show some discrepancies with the correct $D_{g}$ for some DE models and/or $\Omo$ values. Thus, one needs to be very careful when one extends the approximate solutions to the general models and/or other cosmological parameters.

Even though we have obtained the exact analytic solution of $D_{g}$ for the general DE models, this solution is limited to the constant $\omde$ models.  Thus, the applications of this solution to the real observations are very limited. However, we can apply this solution to the more general cases like the time-varying $\omde$ by interpolating between models with constant $\omde$. We have found that $D_{g}$ and $f$ obtained from the constant $\omde$ and the time-varying one are quite different even though we have the same values of $\omde$s at present. If we are able to obtain a good constraint on $\omde$ from the cosmological background evolution observations, then we will be able to constrain $D_{g}$ and $f$ very accurately. Thus, the exact analytic solution of $D_{g}$ can be used as the very useful tool for the interpretation of LSS survey data.

%%%%%%%%%%%%%%%%%%%%%%%%%%%%%%%%%%%%%%%%%%%%%%%%%%%%%%%%%%%%%%%%%%%%%%%%
\section*{Acknowledgments}
%%%%%%%%%%%%%%%%%%%%%%%%%%%%%%%%%%%%%%%%%%%%%%%%%%%%%%%%%%%%%%%%%%%%%%%%%
This work was supported in part by the National Science Council, Taiwan, ROC under the Grants NSC 95-2112-M-001-052-MY3 (KWN) and the National Center for Theoretical Sciences, Taiwan, ROC.

\end{document}